# SPIN WAVES IN A PERIODICALLY LAYERED MAGNETIC NANOWIRE

V.V. Kruglyak and R.J. Hicken

School of Physics, University of Exeter, Stocker Road, Exeter, EX4 4QL, UK

A.N. Kuchko

Donetsk National University, 24 Universitetskaya Street, Donetsk, 83055, Ukraine

V.Yu. Gorobets

Institute of Magnetism of NAS of Ukraine, 36-b Vernadskogo Blvd, 03142, Kyiv, Ukraine

We report a simple theoretical derivation of the spectrum and damping of spin waves in a cylindrical periodically structured magnetic nanowire (cylindrical magnonic crystal) in the "effective medium" approximation. The dependence of the "effective" magnetic parameters upon the individual layer parameters is shown to be different from the arithmetic average over the volume of the superlattice. The formulae that are obtained can be applied firstly in the description of spin wave dispersion in the first allowed band of the structure; and secondly in the design of a magnonic crystal with band-gaps in an arbitrary part of the spin wave spectrum.

## Introduction

Materials with magnetic properties periodically modulated at the nanometer scale (so called magnonic crystals, or magnetic superlattices - MSL) have potential for applications in magneto-electronic devices. For example, a one-dimensional thin film MSL is known to possess properties that cannot be reduced to those of the separate layers. Also, phenomena such as giant magnetoresistance (GMR)[1], large out-of-plane magnetic anisotropy[2], and magnetic field controlled photonic[3] and magnonic[4-6] band gaps have been observed. However, in other situations MSLs can be thought of as "effective media" with "effective" parameters[7] that represent an average of those of the constituent layers. A patterned recording medium consisting of a two-dimensional periodic array of nanosized magnetic elements[8] represents another example of a periodic magnetic structure. Due to the demand for greater recording densities, the element size and separation are continuously decreasing, promoting interactions between elements via stray magnetic fields. Therefore the dynamical response of an individual element is determined by the spectrum of collective excitations of the entire array. Moreover, current experimental methods do not yet provide sufficient spatial resolution for the dynamical properties of a single element to be studied directly. Consequently the magnetic properties of the element must be deduced from measurements made on the entire array.

Arrays of cylindrical magnetic nanowires deposited electrochemically within porous membranes[9] have attracted much attention due to their relative ease of fabrication and because of their potential for use as magneto-photonic crystals and recording media[10]. In the latter case an element density of 1 Tbit/inch$^2$ has recently been reported[11]. Arrays of multilayered nanowires can also be produced by this method[12,13], providing an example of a three-dimensional (3D) MSL. Such structures are important for the field of magneto-photonics, for magnetic recording technology through exploitation of the current perpendicular to the plane (CPP) GMR effect, and in designing patterned media with tunable "effective" magnetic properties. Finally, a 3D MSL may also act as a 3D magnonic crystal, although the strong mag-netic damping present in metallic structures must be addressed if they are to be used as media for spin wave (SW) propagation.

As the speed of operation of magneto-electronic devices containing MSL approaches the gigahertz regime, the dynamical properties of the device are increasingly determined by the SW spectrum of the MSL, which is strongly influenced by the presence of magnonic band gaps. Due to the continuous trend towards device miniaturization, one must consider MSL with very small characteristic dimensions. As we show below, reduction of the period of the MSL pushes the SW band gaps to higher frequencies. The approximate position of the band gaps and the dispersion of the SW modes in the first allowed band, where the SW wavelength is long compared to the period of structural modulation, is then governed by the "effective medium" parameters. It is therefore important to know how the latter depend upon the parameters of the MSL. In this paper we provide an analytical derivation of the SW spectrum and damping in a single periodic multilayer nanowire (cylindrical MSL) in the "effective medium" approximation, and determine the corresponding "effective" material parameters. The first report upon the fabrication of such structures appeared relatively recently, and to the best of our knowledge there have been no experimental reports upon their dynamical properties. However, since the latter can be investigated by ferromagnetic resonance[14], Brillouin light scattering[15], and possibly magneto-optical pump-probe experiments[16], we anticipate that dynamical studies will appear in the near future.

**Theory and Discussion**

Let us consider an infinitely long cylinder of radius $R$ consisting of two different alternating homogeneous layers. The layers have thicknesses $d_1$ and $d_2$, exchange constants $\alpha_1$ and $\alpha_2$, uniaxial anisotropy constants $\beta_1$ and $\beta_2$, Gilbert damping parameters $\lambda_1$ and $\lambda_2$, and gyromagnetic ratios $g_1$ and $g_2$ ($g_j>0$, $j=1,2$). The easy axis (EA) and the external bias magnetic field $\vec{H}_0$ are aligned parallel to the axis of the cylinder, which is also an easy axis for the shape anisotropy. The value of the saturation magnetization $M_0$ is assumed to be constant throughout the entire sample. In the static magnetic state the sample is uniformly magnetized along $\vec{H}_0$. The interfaces are assumed to be sharp and flat, and lie perpendicular to the EA. A cylindrical coordinate system is chosen so that the *OZ* axis is parallel to the EA. Hence, the spatial distribution of the material parameters can be described by

$$\chi(z) = \begin{cases} \chi_1, & z_{2n} < z < z_{2n+1}, \\ \chi_2, & z_{2n-1} < z < z_{2n}, \end{cases} \quad z_{2n} = nd, \; z_{2n+1} = nd + d_1, \tag{1}$$

where $\chi$ is one of the parameters $\alpha, \beta, \lambda, g$, the spatial period of the MSL is given by $d=d_1+d_2$, the coordinate of the *n*th interface is $z_n$, and *n* takes the values $0, \pm 1, \pm 2, \ldots$

To describe small perturbations from the ground state we use the Landau-Lifshitz-Gilbert equations

$$\frac{\partial \vec{M}_j}{\partial t} = -g_j \left[\vec{M}_j \times \vec{H}_{E,j}\right] + \frac{\lambda_j}{M_0}\left[\vec{M}_j \times \frac{\partial \vec{M}_j}{\partial t}\right], \tag{2}$$

where $\vec{M}_j$ is the magnetization. The effective magnetic field $\vec{H}_{E,j}$ is given by

$$\vec{H}_{E,j} = (H_0 + \beta_j M_0)\vec{n} + \frac{\partial}{\partial \vec{r}}\left(\alpha_j \frac{\partial \vec{M}_j}{\partial \vec{r}}\right) + \vec{h}_{m,j}, \tag{3}$$

where $\vec{n}$ is a unit vector along *OZ*, $\vec{h}_{m,j} = -\vec{\nabla}\varphi_j$ is the magneto-dipole field, and the magnetic potential $\varphi_j$ is determined from

$$\nabla^2 \varphi_j - 4\pi \left(\frac{\partial M_{j,x}}{\partial x} + \frac{\partial M_{j,y}}{\partial y}\right) = 0. \tag{4}$$

Following the method described in Ref. 17, we write the magnetization as

$$\vec{M}_j(\vec{r}, t) = \vec{n} M_0 + \vec{m}_j(\vec{r}, t), \tag{5}$$

where $\vec{m}_j$ is a small perturbation from the ground state ($|\vec{m}_j| \ll M_0$). Then, assuming a periodic dependence upon time, the latter becomes $\vec{m}_j(\vec{r}, t) = \vec{m}_j(\vec{r})\exp(-i\omega t)$, and after some algebra one obtains

$$\left[\Omega_j^2 - \left(\tilde{H}_j - \alpha_j \nabla^2\right)\left(\tilde{H}_j + 4\pi - \alpha_j \nabla^2\right)\right]\nabla^2 \varphi_j + 4\pi\left(\tilde{H}_j - \alpha_j \nabla^2\right)\frac{\partial^2 \varphi_j}{\partial z^2} = 0, \tag{6}$$

where $\Omega_j = \omega/g_j M_0$, $\tilde{H}_j = h + \beta - i\Omega_j \lambda_j$, and $h = H_0/M_0$. This equation has solutions of the form

$$\varphi_j = J_m(\kappa_j \rho)\exp\{i(m\psi + G_j z)\}, \tag{7}$$

where $J_m$ is the modified Bessel function of order $m$ ($m$ is an integer), $\rho$ is the distance from the cylinder axis, and $\psi$ is the azimuthal angle. The axial and radial wave numbers $G_j$ and $\kappa_j$ must satisfy the following relation

$$\begin{aligned}\alpha_j^2 \left(G_j^2 + \kappa_j^2\right)^3 + \alpha_j\left(\tilde{B}_j + \tilde{H}_j\right)\left(G_j^2 + \kappa_j^2\right)^2 + \\ + \left(\tilde{H}_j \tilde{B}_j - \Omega_j^2 - 4\pi\alpha_j G_j^2\right)\left(G_j^2 + \kappa_j^2\right) - 4\pi\tilde{H}_j G_j^2 = 0\end{aligned}, \tag{8}$$

where $\tilde{B}_j = \tilde{H}_j + 4\pi$. In the absence of dissipation ($\lambda_j=0$), this equation reduces to that derived in Ref. 17. In the general case, another relation between $\Omega_j$, $m$, $G_j$ and $\kappa_j$ must be found by application of appropriate boundary conditions on the cylinder surface $\rho=R$, but this requires numerical calculations. An analytical solution may be obtained in the limit of a thin nanowire $R<\xi_{ex}$, where $\xi_{ex}$ is the exchange length. In this case, the magnetization can be assumed to be uniform within the cross-section of the cylinder ($\kappa_j=0$)[17,18], and the following dispersion relation for SW in a layer of type $j$ is obtained

$$G_j = \sqrt{\frac{1}{\alpha_j}\left[\Omega_j - h - \beta_j - i\Omega_j \lambda_j\right]}. \tag{9}$$

To find the SW spectrum in the entire sample, we use the Bloch theorem and impose exchange boundary conditions (without interface anisotropy) at the interfaces at $z_n$[19]

$$m_1\big|_{z_n} = m_2\big|_{z_n}, \quad \alpha_1 \frac{\partial m_1}{\partial z}\bigg|_{z_n} = \alpha_2 \frac{\partial m_2}{\partial z}\bigg|_{z_n}, \tag{10}$$

and finally arrive at the following expression for the SW spectrum in a thin cylindrical MSL

$$\cos(Kd) = \cos(G_1 d_1)\cos(G_2 d_2) - \frac{1}{2}\left(\frac{\alpha_2 G_2}{\alpha_1 G_1} + \frac{\alpha_1 G_1}{\alpha_2 G_2}\right)\sin(G_1 d_1)\sin(G_2 d_2), \qquad (11)$$

where $\mathrm{Re}(K) = k$ is the SW quasi-wave number and $\mathrm{Im}(K) = \tilde{\kappa}$ is the inverse of the effective SW attenuation length. This equation is well known since it is identical to that obtained for a SW in a thin film MSL in the short wave approximation[20,21]. This means that the effects predicted by these earlier calculations must also appear in the present case. However, the results discussed below, including expressions for the "effective medium" parameters, can also be applied to a thin film MSL when the exchange dominates.

The spectrum and damping of a SW in the "effective medium" limit are now derived from (11) by assuming that the SW wavelength is greater than the spatial period of the MSL ( $\mathrm{Re}(G_j d) \ll 1$, $\mathrm{Re}(Kd) \ll 1$ ) [20]. In this limit the SW will see the MSL as being quasi-uniform with "effective medium" parameter values corresponding to a certain average over the unit cell of the MSL. We obtain

$$K_{eff} \approx \sqrt{\frac{1}{\bar{\alpha}}\left[\frac{\omega}{M_0 \bar{g}} - h - \bar{\beta} - \frac{\omega \bar{\lambda}}{M_0 \bar{g}}\right]}, \qquad (12)$$

where the "effective medium" parameters are given by

$$\bar{\beta} = \frac{\beta_1 d_1 + \beta_2 d_2}{d_1 + d_2}, \qquad (13)$$

$$\bar{\alpha}^{-1} = \frac{\alpha_1^{-1} d_1 + \alpha_2^{-1} d_2}{d_1 + d_2}, \quad \bar{g}^{-1} = \frac{g_1^{-1} d_1 + g_2^{-1} d_2}{d_1 + d_2}, \qquad (14)$$

$$\bar{\lambda} = \frac{\lambda_1 g_2 d_1 + \lambda_2 g_1 d_2}{g_2 d_1 + g_1 d_2}. \qquad (15)$$

The effective anisotropy constant $\bar{\beta}$ is given by an arithmetic average over the volume of the MSL (13), as one might expect. However, the effective exchange constant $\bar{\alpha}$ and gyromagnetic ratio $\bar{g}$, are instead obtained in (14) from an arithmetic average of the quantities $\alpha_j^{-1}$ and $g_j^{-1}$. As may be seen from Fig. 1, the rule for calculating the effective value dif-fers significantly from the arithmetic average (13). The "effective medium" Gilbert damping parameter is even more complicated (15), depending also upon the values of the gyromagnetic ratio in the layers. The effective anisotropy and exchange parameters have been observed experimentally in thin film MSL's by microwave FMR. We therefore expect that such measurements may also be used for the experimental observation of the effective damping parameter and gyromagnetic ratio. In order to avoid consideration of non-uniform demagnetizing fields[22], we have assumed that the saturation magnetization $M_0$ is not modulated in the structure. However, modulation of $M_0$ may be analytically considered in a thin film MSL, in which case it is easy to see from (9) and (11) that the dependence of the "effective medium" parameters upon those of the layers would be even more complicated.

A uniform saturation magnetization also means that the magnetostatic fields outside the periodic multilayer nanowire are the same as those for a uniform nanowire. Hence, the formalism developed for SWs in an array of uniform nanowires[5,23] could be generalized to describe the dynamics in an array of periodic multilayer nanowires. However, this is beyond the scope of the present paper. In principle, a cylindrical MSL in which the saturation magnetization is constant but other magnetic parameters are modulated, could be made from Co-P alloys[24,25]. The magnetic parameters of the latter are very sensitive to the phosphorous concentration due to transitions from the amorphous to the crystalline state and from a hexagonal to a cubic structure. Parameter values representative of the Co-P system are used in the graphs presented here.

In Fig. 2 we plot the SW spectrum of a cylindrical MSL (11) in which the uniaxial anisotropy is modulated, together with the spectrum of an "effective medium" (12) with the "effective" anisotropy calculated from (13) using the same layer parameters. We see that band gaps emerge at "effective medium" SW frequencies for which

$$K_{eff} = \frac{\pi l}{d}, l=1,2,....  \quad (16)$$

This may be understood by noting that the linearised Landau-Lifshitz equation without damping may be recast in a form identical to that of the Schrödinger equation, in which the periodically modulated anisotropy is analogous to a periodic electronic potential[26]. Modulation of the exchange constant and gyromagnetic ratio also leads to the formation of band gaps, although these quantities are not analogous to the electron potential, and the corresponding band gaps shift towards higher frequencies. Moreover, the width and position of the band gaps has a significant dependence upon the ratio of the MSL layer thicknesses (Fig. 3). Nevertheless, condition (16) still provides a useful starting point for designing a MSL with desired band gap parameters. Further tuning of the MSL parameters can be achieved using the graphical technique described in Ref. 6 and 19. It is interesting to note that the position and the width of the band gap can be tuned independently. Equation (16) shows that the position of the first band gap can be set by adjusting only the period of the MSL, which is probably the easiest parameter of the MSL to change. The width of the band can be varied by changing the magnetic parameters of the layers and the ratio of their thicknesses, while leaving the "effective" parameters (13)-(15) unchanged (Fig. 3).

In Ref. 27 and 28 we considered (in the short wave approximation) SW in a thin film periodic multilayer with interfaces of finite thickness, but without taking into account modulation of the gyromagnetic ratio. Those models are easily applied to a cylindrical MSL, and the results derived previously remain valid. On the other hand, expressions (9), (11)-(15), which are derived here after the inclusion of a periodically modulated gyromagnetic ratio, are easily generalized to those other models.

## Summary

The spectrum and damping of SW in a cylindrical MSL have been derived in the "effective medium" approximation. It has been shown that the "effective medium" parameters have an unexpected dependence upon those of the constituent layers and, in general, are not given by an arithmetic average over the volume of the MSL. The formulae that have been obtained are useful in designing composite magnetic materials with desired microwave properties.

## Acknowledgments

The authors thank Prof. Yu.I. Gorobets for fruitful discussions.

10
List of figure captions

Figure 1   Ratio $S$ of an "effective medium" parameter $\chi$ given by (14) to that given by (13) is plotted for the same values of the parameter in the layers $\chi_i$.

Figure 2   The spectrum of SW in a cylindrical MSL (11) with modulation of the constant of uniaxial anisotropy $\beta_2/\beta_1 = 5$

1. d = 5 nm,

2. d = 10 nm,

3. d = 20 cm.

The dashed line represents the SW spectrum in a fine layered nanowire (12).

Figure 3   The spectrum of SW in a cylindrical MSL (11) with modulation of the constant of a) the exchange parameter $\alpha_2/\alpha_1 = 5$, and b) the gyromagnetic ratio $g_2/g_1 = 5$. In both cases d = 20 nm and

1.   $d_2/d_1 = 2$,

2.   $d_2/d_1 = 5$,

3.   $d_2/d_1 = 10$.

The dashed lines represent the SW spectrum in a fine layered nanowire (12).

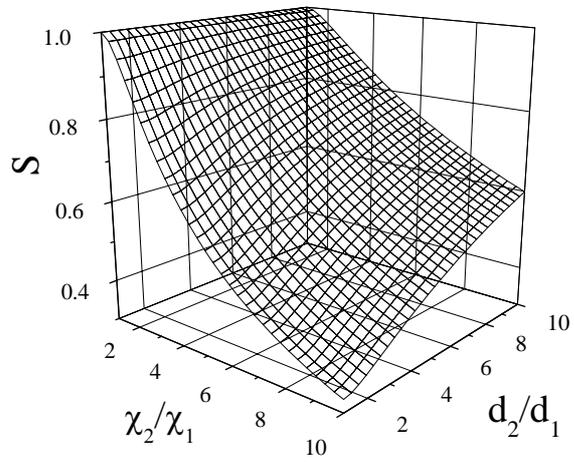

Fig. 1

Kruglyak et al

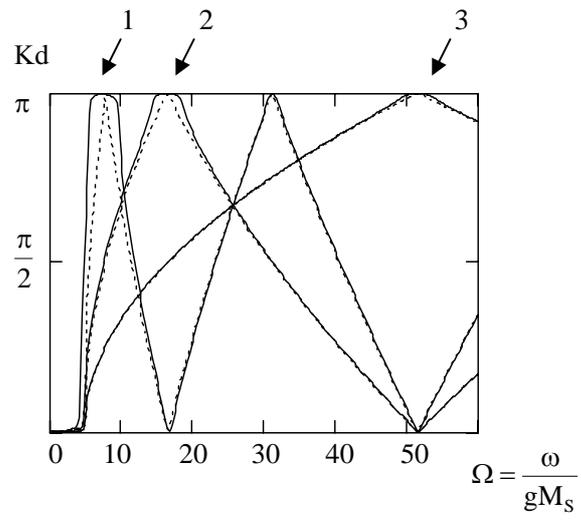

Fig. 2

Kruglyak et al

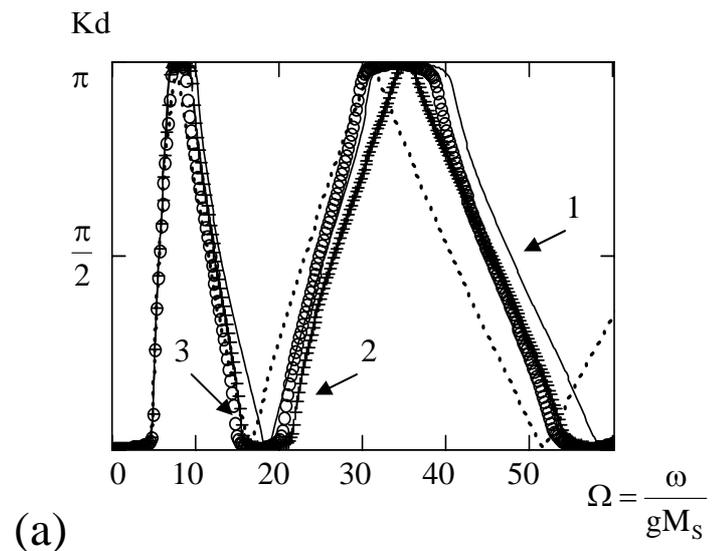

(a)

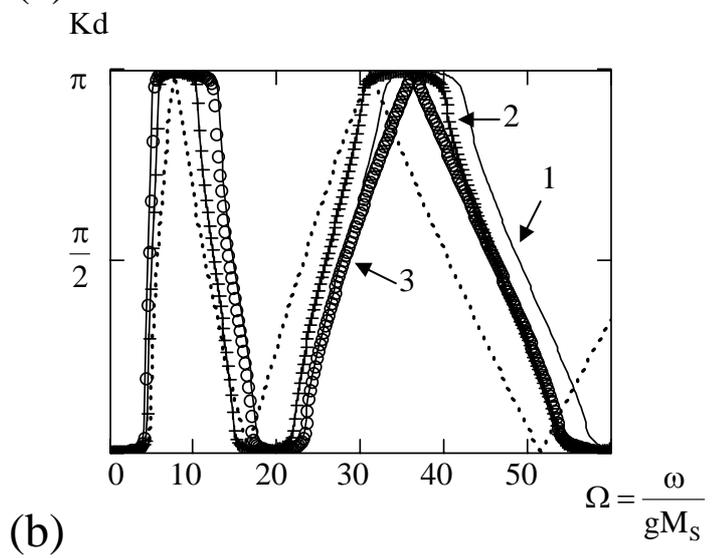

(b)

(

Fig. 3

Kruglyak et al